\documentclass[sigconf,nonacm]{acmart}

\AtBeginDocument{%
  \providecommand\BibTeX{{%
    \normalfont B\kern-0.5em{\scshape i\kern-0.25em b}\kern-0.8em\TeX}}}

\usepackage{cleveref}

\newcommand{\todo}[1]{}
\renewcommand{\todo}[1]{{\color{red}[[TODO: {#1}]]}}

\begin{document}

\title[The OCEAN mailing list data set]{The OCEAN mailing list data set: Network analysis spanning mailing lists and code repositories}


\author{Melanie Warrick}
\email{warrick.melanie@gmail.com}
\affiliation{%
  \institution{Google, Inc.}
  \city{Mountain View}
  \state{California}
  \country{USA}
  \postcode{94043}
}

\author{Samuel F. Rosenblatt}
\email{samuel.f.rosenblatt@uvm.edu}
\orcid{0000-0002-9249-6423}
\affiliation{%
  \institution{University of Vermont}
  \streetaddress{82 University Pl.}
  \city{Burlington}
  \state{Vermont}
  \country{USA}
  \postcode{05405}
}

\author{Jean-Gabriel Young}
\email{jean-gabriel.young@uvm.edu}
\orcid{0000-0002-4464-2692}
\affiliation{%
  \institution{University of Vermont}
  \streetaddress{82 University Pl.}
  \city{Burlington}
  \state{Vermont}
  \country{USA}
  \postcode{05405}
}

\author{Amanda Casari}
\email{amcasari@google.com}
\orcid{0000-0002-3201-8964}
\affiliation{%
  \institution{Google, Inc.}
  \city{Mountain View}
  \state{California}
  \country{USA}
  \postcode{94043}
}

\author{Laurent H\'ebert-Dufresne}
\email{laurent.hebert-dufresne@uvm.edu}
\orcid{0000-0002-0008-3673}
\affiliation{%
  \institution{University of Vermont}
  \streetaddress{82 University Pl.}
  \city{Burlington}
  \state{Vermont}
  \country{USA}
  \postcode{05405}
}

\author{James Bagrow}
\email{james.bagrow@uvm.edu}
\orcid{0000-0002-4614-0792}
\affiliation{%
  \institution{University of Vermont}
  \streetaddress{82 University Pl.}
  \city{Burlington}
  \state{Vermont}
  \country{USA}
  \postcode{05405}
}

\renewcommand{\shortauthors}{Warrick et al.}

\begin{abstract}
Communication surrounding the development of an open source project largely occurs outside the software repository itself. Historically, large communities often used a collection of mailing lists to discuss the different aspects of their projects. Multimodal tool use, with software development and communication happening on different channels, complicates the study of open source projects as a sociotechnical system. Here, we combine and standardize mailing lists of the Python community, resulting in 954,287 messages from 1995 to the present. We share all scraping and cleaning code to facilitate reproduction of this work, as well as smaller datasets for the Golang (122,721 messages), Angular (20,041 messages) and Node.js (12,514 messages) communities. To showcase the usefulness of these data, we focus on the CPython repository and merge the technical layer (which GitHub account works on what file and with whom) with the social layer (messages from unique email addresses) by identifying 33\% of GitHub contributors in the mailing list data. We then explore correlations between the valence of social messaging and the structure of the collaboration network. We discuss how these data provide a laboratory to test theories from standard organizational science in large open source projects.
\end{abstract}

\begin{teaserfigure}
  \centering
  \includegraphics[width=0.77\textwidth]{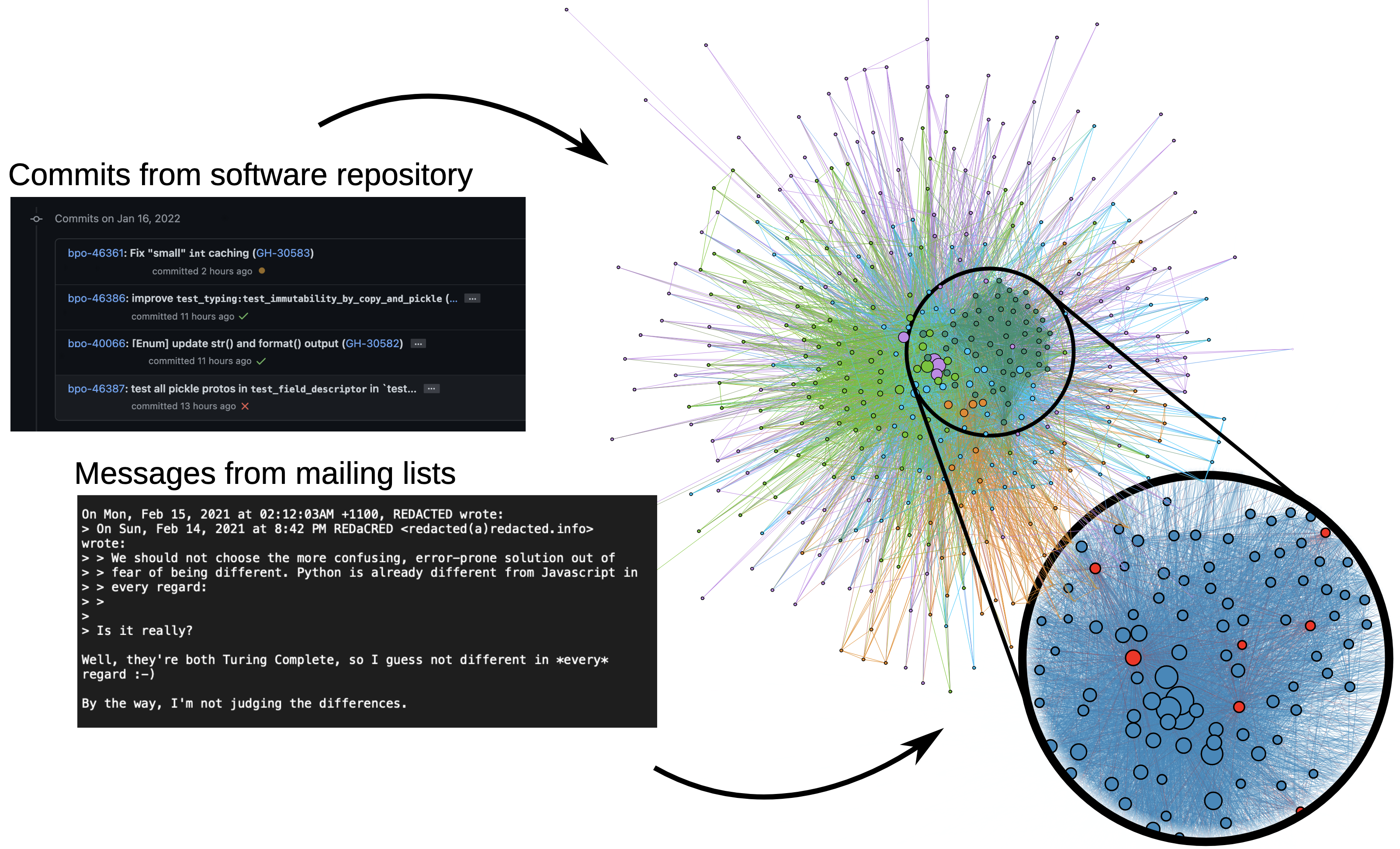}
  \caption{\textbf{Layering social communication data onto the technical network structure of an open source project.} The main network is produced by projecting the account-commit-file data of the CPython repository on a ``graph of who collaborates with whom'' which unveils the modular structure of the project and community (nodes are colored to highlight this using Louvain modularity maximization \cite{blondel2008fast}). We then zoom on a particular module to highlight the role of certain nodes in the social layer of the community using our mailing list data and find that 6 of the top 10 nodes ranked by counts of toxic words are found in this single module.}
  \label{fig:teaser}
\end{teaserfigure}

\maketitle

\section{Introduction}
The practice of mining and analysing software repositories provides a unique window into software development and the social practice of problem solving as a whole. Unfortunately, these analyses often come with the caveat that much of the collaborative work other than coding appears in public but away from the software repository itself~\cite{eghbal2020working}.  The community relies on data streams that, unlike \texttt{git} history for example, do not come in standard formats~\cite{hassan2008road}: bug reports, mailing lists, online forums, etc.

For large open source communities (OSC), social interactions are rarely limited to a single platform~\cite{wiggins2008social}, from direct messages, to targeted forums or massively distributed mailing lists. Importantly, discussions there concern a much wider range of contributions to the community than coding itself, with efforts around the development of norm for the community or organisation of social events~\cite{raymond1999cathedral,Young2021which,}. These interactions have been shown to be critical to the growth and health of the community, with constructive and timely replies being positively correlated with future participation~\cite{Jensen2011joining}.

Data describing social interactions within OSCs are notoriously hard to analyse given their unstructured format~\cite{Rigby2007what,hassan2008road}. Past work has therefore focused on manual readings of samples of messages, for example from newcomers \cite{Jensen2011joining} or major developers and community members  \cite{Rigby2007what}, or on automated analysis of specific windows chosen for their fixed format \cite{Dam2015mining}. Altogether, several challenges exist to mining mailing lists \cite{Bird2006mining,guzzi2013communication}. Here, we present data with scraping software to address two specific problems: Large communities tend to use multiple mailing lists for different topics (e.g. development, ideas, announcements, etc.) and these different lists do not operate with the same software, encoding, or data structure.

\begin{figure*}[th!]
    \centering
    \includegraphics[width=0.95\linewidth]{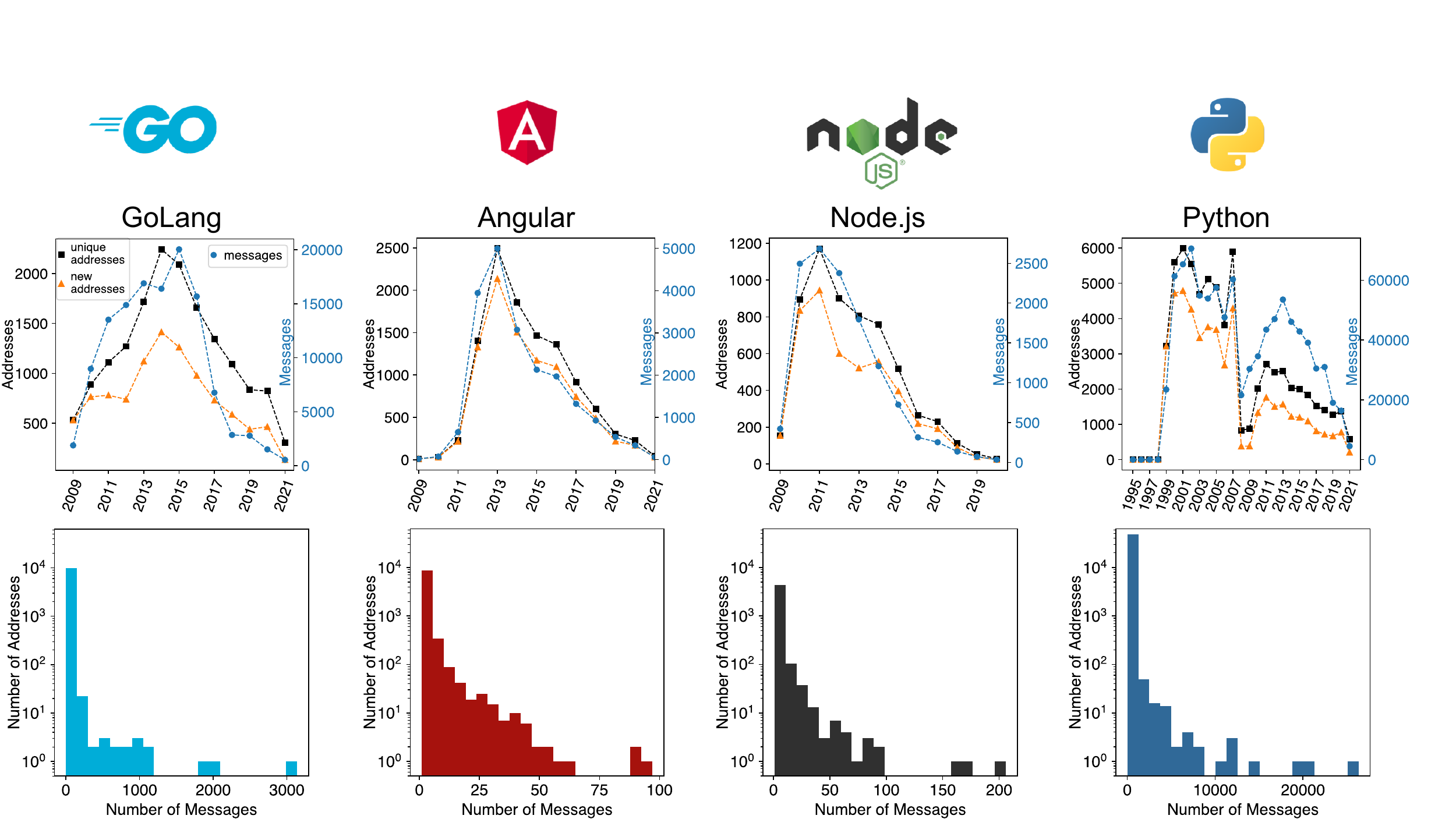}
    \caption{(top row) Volumes of addresses and messages of different mailing list datasets through time. Messages and addresses are on different scales. (bottom row) Histograms of messages per address. 
    Three addresses with outlying volumes of automated messages (7,592,  19,492, and 33,660) are not shown in the GoLang histogram (lower left).}
    \label{fig:histograms}
\end{figure*}

\section{The OCEAN mailing list data}

We present the data collected as part of the Open-source Complex Ecosystem And Networks (OCEAN) partnership between Google Open Source and the University of Vermont. Our code base for data collection is hosted at \url{https://github.com/google/project-OCEAN} and contains a suite of utilities to compile mailing lists based on their hosting service. This is currently limited to mailing lists hosted through Mailman, Google Groups, or Pipermail.

We include 14 mailing lists as part of this initial release but the data set could be straightforwardly extended to include other OSCs which use mailing lists as a main form of communication. We here show all lists currently available, in chronological order:
\begin{enumerate}
    \item \texttt{pipermail-python-dev}, started on 1995-03,
    \item \texttt{pipermail-python-list}, started on 1999-02,
    \item \texttt{pipermail-python-announce-list}, started on 1999-04,
    \item \texttt{mailman-python-announce-list}, started on 1999-04, 
    \item \texttt{mailman-python-dev}, started on 1999-04,
    \item \texttt{mailman-python-ideas}, started on 2006-12,
    \item \texttt{pipermail-python-ideas}, started on 2006-12, 
    \item \texttt{gg-nodejs}, started on 2009-06, 
    \item \texttt{gg-angular}, started on 2009-09, 
    \item \texttt{gg-golang-checkins}, started on 2009-11,
    \item \texttt{gg-golang-dev}, started on 2009-11, 
    \item \texttt{gg-golang-nuts}, started on 2009-11,  
    \item \texttt{gg-golang-announce}, started on 2011-05, 
    \item \texttt{gg-golang-codereviews}, started on 2013-12.
\end{enumerate}
This database was aggregated by consulting with open source community members to identify both active mailing lists used by the selected projects, as well as mailing lists which are known to contain historical decision making information.

Our codebase then standardizes date format, text information and origins of messages under one large database per community. The most important fields of our data sets are as follows:

\begin{center}
\begin{tabular}{||c c c||} 
 \hline
 Field & Type & Example \\ [0.5ex] 
 \hline\hline
 from\_name & string & Example Person \\ 
 \hline
 from\_email & string & example.person@gmail.con \\
 \hline
 to\_name & string & python-dev \\
 \hline
 to\_email & string & python-dev@python.org  \\
 \hline
 message\_id & string & <cc76ef2\ ...\ @googlegroups.com> \\
 \hline
 subject & string & parameters not working  \\
 \hline
 date & datetime & 2016-05-02T11:45:35  \\
 \hline
 body\_text & string & I'm trying to\ldots  \\
 \hline
 original\_url & string & https://groups.google.com/\ldots \\
 \hline
\end{tabular}
\end{center}

In the earlier years of the archives, the date format used by the mail clients was quite variable. While our tool supports a series of queries to parse dates, ``to,'' ``from,'' and references fields,  it does not always succeed, so we also preserved the raw data under additional fields to allow easy format checks.

Key descriptive statistics of the mailing lists data are shown in Fig.~\ref{fig:histograms}. In terms of volume of both messages and newcomers, all mailings lists most recently peaked between 2011 and 2015. Across all communities, the ratio of newcomers to established addresses tend to be relatively fixed but varied with some mailing lists driven by newcomers (Angular) or existing addresses (GoLang). In all cases, we find skewed distributions of messages per unique email address, with outliers most often representing bots. These bots can be easily identified within the data, but are preserved as they often forward actual social messages from other platforms.
    
\section{Current use of the database}

The OCEAN mailing list database was recently shared with researchers at Galois, inc. as part of a DARPA funded project on the impact of toxicity on collaborations in large and important open source projects. As part of this project, we here focus on a subset of the Python community: the CPython repository containing the reference implementation of the language and where most Python development occurs. The database was leveraged as one element in a new tool, the LAGOON platform for Leveraging AI to Guard Online Open-source Networks, which does identity merging and comes packaged with a UI for inspecting and analyzing the ingested open source community \cite{LAGOON}. Through this tool, our collaboration highlights that 33.2\% of all CPython contributors can be identified in the mailing list data. This provides us with a large sampling of the content and valence of discussions within the community.

As a simple case study, we can tag CPython contributors by the amount of toxic language they send on the mailing list.
Previous research in management science has explored the
positive association between power and negative ties~\cite{labianca2006}, and
it is unclear if these results generalize to OSS communities which lack clear hierarchies. 
For simplicity, we here study this issue with a naive definition of toxic language, based on a previous study which developed an annotated corpus and lexicon for harassment research \cite{rezvan2018quality}. This corpus was manually edited to account for words which tend to have a different meaning in technical discussions than in colloquial language (e.g. ``primitive'').
We then summarize the social messages of a given users simply by counting how many toxic words they have used while messaging the community.

In Fig.~\ref{fig:teaser}, we show where toxic nodes are found within the collaboration network of CPython, a projection of the account-commit-files network on the set of edges of accounts that have touched files in common). To do so, we first identify modules (e.g. teams) in the community using the Louvain modularity maximization algorithm \cite{blondel2008fast}. 
We then find that toxic language is not randomly distributed within the collaboration network of CPython, with 6 of the top 10 nodes ranked by counts of toxic words found within a single central module.

\begin{figure*}[th!]
    \centering
    \includegraphics[width=0.95\linewidth]{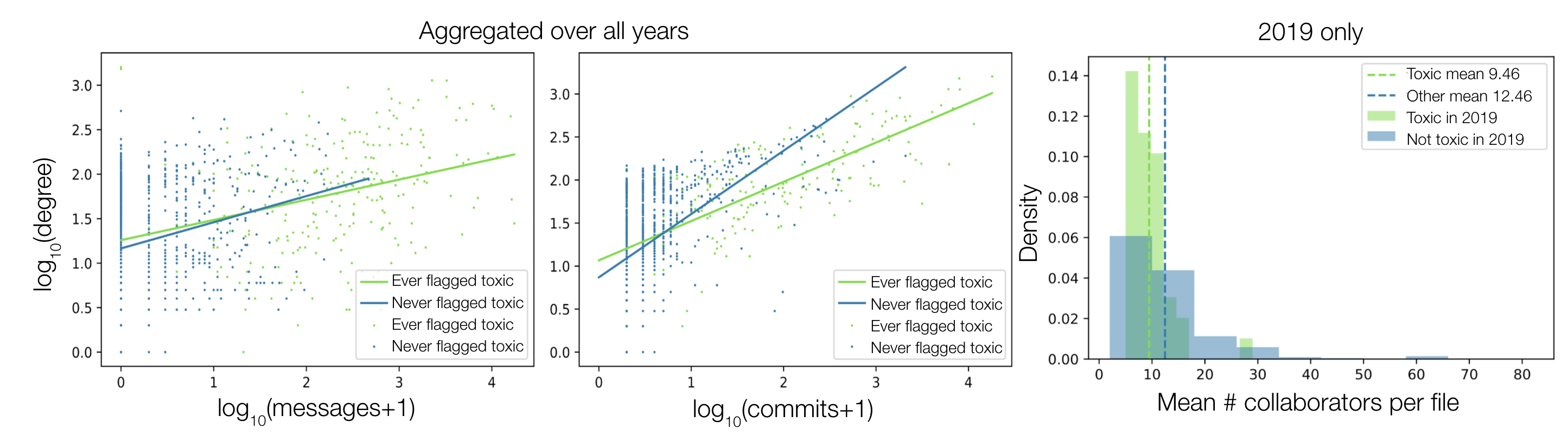}
    \caption{Correlations between activity in the social (messages and toxicity) and technical (commits, collaborators or ``degree'', and files) layers of the CPython repository. Degree refers to a user's number of co-editors in the projected version of the user-commit-file network. Accounts that use toxic words tend to have less collaborators per commit and per file.}
    \label{fig:correlations}
\end{figure*}

In Fig.~\ref{fig:correlations}, we quantify the correlations between toxic word usage and network structure in the collaboration network. 
In the right panel we first find, unsurprisingly, that 
(1) accounts that send more messages are more likely to be flagged as toxic (more opportunity to use toxic words) and that 
(2) regardless of toxicity, accounts that send more messages also tend to have more collaborators in the technical layer of the network. 
More surprisingly, in the middle panel, we find that 
(3) toxic accounts tend to gain less collaborators for every commit to the repository. This result is further explored in the right panel where we see that accounts that are toxic in a given year tend to have a smaller average number of collaborators per file.

Further research is needed to fully understand these results and, in particular, to delineate between toxic accounts that as contributors are central technical dependencies and toxic accounts that operate on the periphery of a project and offer minimal contributions.
Many different mechanisms can be postulated to explain the observed correlations: perhaps having few collaborators lead to toxic languages, or conversely, toxic languages lead to disengagement of collaborators. Altogether, these simple correlations showcase the value of merging social communication data like mailing lists with technical data from the repository itself, but warrant further research to disentangle the mechanisms at play.

\section{Additional related work}
There are several data sets aiming to capture activities around software development outside of its code base.
Besides the analyses of mailing lists already mentioned~\cite{Bird2006mining,guzzi2013communication,wagstrom2005social,ducheneaut2005socialization}, relevant studies includes discussions on unusual platforms such as blogs~\cite{wagstrom2005social}, Slack messages~\cite{chatterjee2019exploratory,chatterjee2020software} or IRC messages~\cite{yu2011communications,shihab2009studying}, though the data for these studies is not always made available in easy to parse format, if at all.
Going further afield, datasets of Stack Overflow messages and events~\cite{MSRChallenge2013,MSRChallenge2015,msr2019challenge} are also related to our work in that they pertain to a community and are disconnected from the code base.
However, discussions on Stack Overflow focus on using the software rather than developing it~\cite{manes2019often}, and thus capture a different aspect of a community's behavior.

Our example application---natural language processing and sentiment analysis of data pertaining to open source communities---is also a very active field~\cite{calefato2018sentiment}.
Work in this area uses diverse data source such as commit message~\cite{sinha2016analyzing}, issue reports~\cite{mantyla2016mining,mantyla2017bootstrapping}, email lists~\cite{Rigby2007what,rousinopoulos2014sentiment}, or Q\&A platform, to name a few~\cite{calefato2018sentiment}.
The main challenges here is the unusual nature of technical communication, which differs significantly from the type of text usually analyzed with NLP~\cite{mantyla2018natural}.

\section{Challenges, improvements, and future work}
From the projects included in our data repository, we find the practice of using mailing lists as the main communication channel for open source projects has been steadily decreasing in popularity over the last decade or so. This presents an obvious limitation of relying solely on mailing lists as a proxy for social interactions within a community, but also an provides an archival opportunity as our database creates a largely static window into the past social interactions around current major projects in open source software development. For continued study of active projects within these communities, our data will need to be supplemented by forum discussions, bug reports, and newer platform conversation channels such as messaging platforms, social media, and blog rolls~\cite{wagstrom2005social,fang2020need}.

In addition to the raw text data associated with messages, we expect to process the mailing lists to automatically identify text structures such as email signatures, code and quotes in replies~\cite{rahman2019cleaning}. Quotes, in particular, are potentially powerful as they provide a window into moderation norms and practices. Indeed, targeted toxic emails are often moderated and do not appear as a sent message in the mailing list, but targeted recipients can still receive this message directly rather than through the list. For example, by sending the email to both \texttt{target\@email.com} and to the mailing list, the target receives the email regardless of moderation. The target can then reply to the message, which will be quoted in the reply therefore bypassing the original moderation.
Identifying and flagging these quotes will open interesting avenues of research as moderated texts are virtually never included in available data.

A key point of discussion among the project team was how to address personal identifiable information that is already accessible from the data sources, available under the platform terms and conditions, and shared following the community guidelines. Best practices for aggregating, sharing, and working with open source data gleaned from community projects is still mixed~\cite{gold2022ethics}; while transparency is valued in open source, communities also value their personal privacy while working in open spaces. Since we are sharing this database as both a dataset, the list of sources this data was aggregated from, and the source code used to produce it, the team chose to follow the CHAOSS Community Data Policy, which states that ``our community data is part of our public history,'' disclosing all data as assembled from the original sources to ``preserve the authenticity of [our] community data''~\cite{chaoss2022datapolicy}.

Our repository for data compiled using the OCEAN tool will grow and allow the public history of an OSC to be studied in order to evaluate and improve the effectiveness of community norms. Future work will integrate the raw mailing list data with natural language processing tools to classify messages and evaluate how communication, miscommunication, and toxicity can affect collaborations. We then envision studying how the social structure found in the mailing list data reflects the collaborative structure found in the software repository, as well as the architecture of the software produced; from toxic messages and drive-by contributions, to social support and long-term innovations.

\begin{acks}
This material is based upon work supported by Google Open Source under the
Open Source Complex Ecosystems And Networks (OCEAN)
project and by the Defense Advanced Research Projects Agency (DARPA) under Contract No. HR00112190092. Any opinions, findings and conclusions or recommendations expressed in this material are those of the author(s) and do not necessarily reflect the views of Google Open Source or DARPA.
\end{acks}

\section*{Author contributions}
M.W. built the data set. S.F.R. ran the analyses. J.G.Y., A.C., L.H.D. and J.P.B. wrote the original draft, which all authors edited. Data are hosted at: \url{https://doi.org/10.6084/m9.figshare.19082540.v2} \cite{Warrick2022}.

\end{document}